\documentclass[aps,twocolumn,twoside,groupedaddress,nofootinbib,nobalancelastpage,
nobibnotes]{revtex4}
\pdfoutput=1
\usepackage[letterpaper, portrait,margin=.6in]{geometry}
\usepackage[hyperfootnotes=false]{hyperref}
\usepackage{amsmath,amsfonts,amssymb,mathrsfs,graphicx,color}
\usepackage[squaren]{SIunits}
\usepackage{verbatim}
\usepackage{enumerate}
\usepackage{graphicx}
\usepackage{xcolor}
\usepackage{slashed}
\usepackage[utf8]{inputenc}
\usepackage{natbib}
\usepackage{multirow}
\usepackage{float}
\usepackage{tabularx}
\usepackage[titletoc,title]{appendix}
\usepackage{graphicx}
\usepackage{amstext}
\usepackage{booktabs}

\newcommand{\ie}{{\it i.e.}}
\newcommand{\eg}{{\it e.g.}}

\newcommand{\cf}{{\it cf.}}

\newcommand{\fig}{Fig.}
\newcommand{\Fig}{Fig.}

\newcommand{\Figs}{Figs.}

\newcommand{\Refer}{Ref.} 

\newcommand{\Refs}{Refs.}

\newcommand{\Tab}{Tab.}

\newcommand{\figu}[1]{\fig~\ref{fig:#1}}
\newcommand{\tabl}[1]{\Tab~\ref{tab:#1}}

\begin{document}

\title{AGN jets as the origin of UHECRs \\ and perspectives for the detection of astrophysical source neutrinos at EeV energies}

\author{Xavier Rodrigues}
\email{xavier.rodrigues@desy.de}
\affiliation{DESY, Platanenallee 6, 15738 Zeuthen, Germany}

\author{Jonas Heinze}
\affiliation{DESY, Platanenallee 6, 15738 Zeuthen, Germany}

\author{Andrea Palladino}
\affiliation{DESY, Platanenallee 6, 15738 Zeuthen, Germany}

\author{Arjen van Vliet}
\affiliation{DESY, Platanenallee 6, 15738 Zeuthen, Germany}

\author{Walter Winter}
\affiliation{DESY, Platanenallee 6, 15738 Zeuthen, Germany}

\date{\today}

\begin{abstract}   
  We demonstrate that a population of Active Galactic Nuclei (AGN) can describe the observed spectrum of ultra-high-energy cosmic rays (UHECRs) at and above the ankle, and that the dominant contribution comes from low-luminosity BL Lacs. 
  An additional, subdominant contribution from high-luminosity AGN is needed to improve the description of the composition observables, leading to a substantial neutrino flux that peaks at EeV energies. We also find that different properties for the low- and high-luminosity AGN populations are required; a possibly similar baryonic loading can already be excluded from current IceCube observations.
  We also show that the flux of neutrinos emitted from within the sources should outshine the cosmogenic neutrinos produced during the propagation of UHECRs. This result has profound implications for the ultra-high-energy ($\sim$EeV) neutrino experiments, since additional search strategies can be used for source neutrinos compared to cosmogenic neutrinos, such as stacking searches, flare analyses, and multi-messenger follow-ups.
  \end{abstract}
  
  \maketitle 
  
  
  Blazars are Active Galactic Nuclei (AGN) with their jets pointed towards Earth; they contribute more than 80\% of the Extragalactic Gamma-ray Background (EGB)~\citep{TheFermi-LAT:2015ykq}, dominating the $\gamma$-ray emission above 50 GeV. In addition, there are strong indications for correlations between the arrival directions of UHECRs and extragalactic $\gamma$-ray sources, including AGN~\citep{Aab:2018chp}. Jetted AGN are also one of the candidate source classes which may have sufficient power to maintain the UHECR flux. As a consequence, it is natural to consider jetted AGN as possible origin of the observed UHECRs. 
  
  In addition, a diffuse flux of high-energy astrophysical neutrinos has been discovered~\citep{Aartsen:2013jdh}. This may be a direct indicator for the origin of UHECRs because neutrinos point back directly to their sources, while UHECRs are deflected by Galactic and extragalactic magnetic fields.  The recent detection of neutrinos from the flaring blazar TXS 0506+056 provides further evidence that cosmic rays are accelerated in AGN up to at least $\sim$PeV energies~\citep{IceCube:2018dnn,IceCube:2018cha}, see also earlier results~\citep{Kadler:2016ygj}. On the other hand, dedicated catalog searches for neutrinos from known AGN limit the contribution of these objects to about 20\% of the diffuse IceCube flux~\citep{Aartsen:2016lir}, which means that resolved $\gamma$-ray blazars are probably not the dominant source of neutrinos at TeV-PeV energies. However, low-luminosity or high-redshift AGN are more numerous and less constrained observationally, and may still, under certain conditions, be a viable candidate for the diffuse IceCube neutrino flux \citep{Palladino:2018lov,Neronov:2018wuo}.
  
  On the other hand, if AGN significantly contribute to the observed UHECR flux, they need to accelerate cosmic-ray nuclei up to $\sim10^{20}$~eV. The production of UHECRs of these extreme energies in AGN is supported by studies involving the simulation of different reacceleration mechanisms \citep{Kimura:2017ubz,Matthews:2018rpe,Mbarek:2019glq}. Photointeractions of these cosmic rays lead to the emission of neutrinos with energies $E_\nu\simeq 0.05 \, E_\mathrm{CR}$/A, where A is the mass number of the cosmic-ray nucleus. AGN would then be expected to yield significant neutrino fluxes in the EeV energy range, where currently only upper limits exist~\citep{Aartsen:2018vtx,Aab:2019auo}.
  
  The production of UHECRs and neutrinos in AGN has been studied in previous works~\citep[\eg][]{Protheroe:1992qs,Mannheim:1993jg,Gorbunov:2002hk,Halzen:2002pg,Dermer:2008cy,Essey:2009ju,Dermer:2010iz,Murase:2011cy,Murase:2014foa,Resconi:2016ggj,Padovani:2016wwn,Rodrigues:2017fmu,Eichmann:2017iyr,Righi:2020ufi}. Other more phenomenology-driven works have had the objective of describing the observed UHECR spectrum and composition with a population of high-energy sources \citep{Fang:2017zjf,Biehl:2017hnb,Boncioli:2018lrv}. However, a self-consistent description of the UHECR spectrum and composition including a neutrino flux prediction from the entire AGN population has not yet been performed.
  
  The objective of this study is two-fold: \textit{(a)} to investigate under what conditions the AGN population can explain the UHECR spectrum observed by the Pierre Auger Observatory (henceforth, \textit{Auger}) \citep{Fenu:2017hlc}, while obeying the most recent IceCube limits at~$\sim$PeV energies~\citep{Aartsen:2016lir}; and \textit{(b)} to investigate the corresponding neutrino fluxes, particularly in the EeV range. This includes both source neutrinos as well as cosmogenic, \ie~those produced in UHECR interactions during their propagation over extragalactic distances.

  Cosmogenic neutrinos are the main target of radio-detection neutrino experiments in the EeV range such as the radio array of IceCube-Gen2~\citep{Aartsen:2019swn}, GRAND~\citep{Alvarez-Muniz:2018bhp}, ARA~\citep{Allison:2011wk} and ARIANNA~\citep{Barwick:2014pca}. Recent descriptions of the UHECR spectrum and composition, however, indicate that the maximum energies are limited by the accelerators, and that they can be described by a rigidity dependence $E_{\mathrm{CR}} \propto Z$, with $Z$ the charge of the nucleus~\citep{Aab:2016zth}. Such a rigidity dependence is generated, for instance, if the particles are magnetically confined to a certain zone of fixed size. This framework leads to low cosmogenic neutrino fluxes~\citep{AlvesBatista:2018zui, Heinze:2019jou}, except for a potential contribution from a subdominant proton population~\citep{vanVliet:2019nse}. 
  By considering AGN as the sources of the UHECRs, we will actually reevaluate the hypothesis that the EeV neutrino sky is dominated by cosmogenic neutrinos, and we will show that source neutrinos may actually be the foreground signal at these energies.


  We assume that cosmic rays are reaccelerated in AGN jets to a power-law spectrum up to ultra-high energies. We then utilize a combined source-propagation model, which has essentially three components (see appendix A for details): \textit{(1)} Simulation of the photohadronic interactions that the accelerated UHECRs undergo in a radiation zone in the jet. This step is relevant in high-luminosity sources, where these interactions lead to efficient neutrino emission and the production of a nuclear cascade, which is simulated self-consistently. \textit{(2)} Propagation of the escaping cosmic rays towards Earth. This also involves the numerical calculation of photointeractions, including the development of a nuclear cascade and the production of cosmogenic neutrinos. \textit{(3)} Extension of the calculation to the entire cosmological distribution of AGN. The resulting overall UHECR and neutrino fluxes are then compared to current measurements.

\begin{figure}[tbp]
  \includegraphics[scale=0.8]{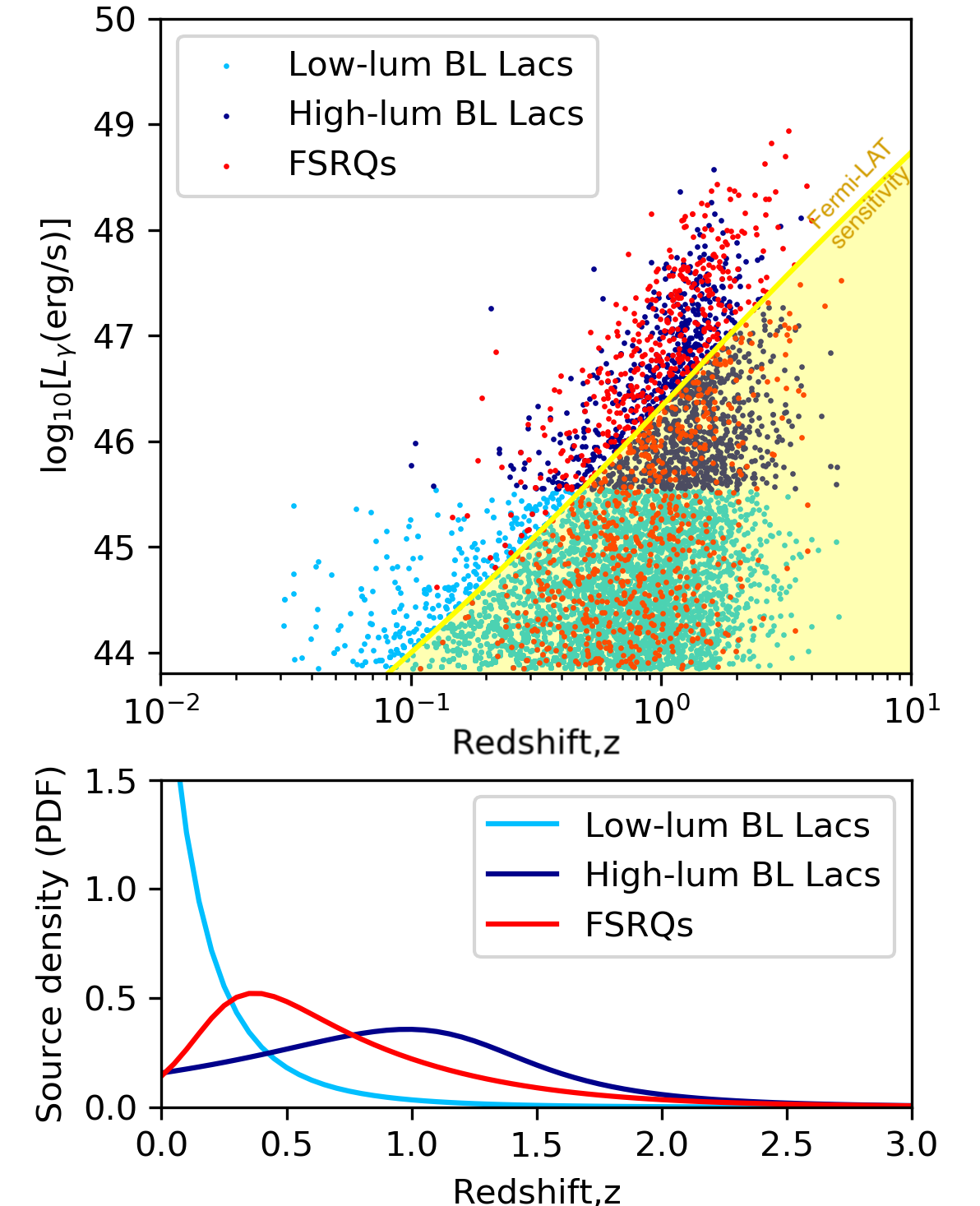}
  \caption{
      Representation of the blazar population as a function of redshift and luminosity, following the model by \citet{Ajello:2011zi, Ajello:2013lka}. The yellow region represents the phase space that falls below the sensitivity of the \textit{Fermi}-LAT $\gamma$-ray telescope.   Here we divide blazars into FSRQs, low-luminosity BL Lacs, and high-luminosity BL Lacs. The lower panel shows the same distribution in redshift (integrated over luminosity) and it clearly illustrates the strong negative evolution of low-luminosity BL Lacs compared to high-luminosity AGN (BL Lacs and FSRQs).
      }
  \label{fig:blazardistro}
  \end{figure}

  We divide AGN into three subpopulations based on their different cosmological evolution functions (\cf~\figu{blazardistro}): low-luminosity BL Lacs, high-luminosity BL Lacs, and flat-spectrum radio quasars (FSRQs); we refer to the latter two categories as ``high-luminosity AGN''. While blazars emit highly beamed $\gamma$-rays and neutrinos in the direction of Earth, the observed UHECRs can originate in the broader family of jetted AGN, since these particles are deflected in magnetic fields. We assume that the cosmic evolution of blazars is representative of this overall AGN population, in view of the unified scheme of radio-loud AGN \citep{Urry:1995mg}.
  We adopt the cosmological evolution model illustrated in \figu{blazardistro} as a function of redshift and $\gamma$-ray luminosity for these AGN subclasses \citep{Ajello:2011zi, Ajello:2013lka}. This evolution model yields a total $\gamma$-ray emissivity from AGN that is consistent with the diffuse \textit{Fermi}-LAT $\gamma$-ray measurements \citep{Abdo:2010nz}.
  Our starting hypothesis is that these AGN populations have similar properties regarding the UHECR acceleration; we will demonstrate, however, that at least the baryonic loading (\ie~the ratio between the amount of energy injected into cosmic rays and that injected into electrons) has to differ for low- and high-luminosity sources in order to satisfy PeV neutrino constraints.
  
  Regarding the chemical injection composition of cosmic rays, we consider a mixture of four representative mass groups: protons, helium-4, nitrogen-14 and iron-56. Their relative abundances, after acceleration and before undergoing photointeractions, follow the composition suggested in \Refer~\citep{Mbarek:2019glq}, namely relative abundances of 1.00, 0.46, 0.30, and 0.14, respectively.
  These fractions correspond to the Galactic cosmic-ray composition and can in fact originate from solar system abundances through chemical enhancement during the acceleration process \citep{Caprioli:2017oun}. 
  Fine-tuning the assumed composition of the primary cosmic rays could of course improve the description of the observed UHECR data, at the cost of extra parameters; however, such detailed description is not the goal of this study. Furthermore, as discussed in appendix A, the conclusions regarding neutrino emission are not sensitive to this particular choice.
  
  The maximum energies of the cosmic rays are determined self-consistently depending on the specific nuclear isotope, based on the balance between the particle's energy loss and acceleration timescales.
  The spectrum of non-thermal photons in the jet is adopted from the \textit{blazar sequence} paradigm \citep{Fossati:1998zn, Ghisellini:2017ico}, an assumption used in previous multi-messenger studies of blazars \citep{Murase:2014foa, Rodrigues:2017fmu}. In this approximation, the non-thermal photon spectrum depends only on the $\gamma$-ray luminosity $L_\gamma$ of the AGN.
  
  The accelerated UHECR nuclei interact with target photons in the AGN jet, leading to photodisintegration and photopion production, which implies the emission of neutrinos with energies roughly following  that of the primary cosmic rays. We simulate these radiative processes including the nuclear cascade in the source  self-consistently, using the numerical code \textsc{NeuCosmA} \citep{Hummer:2010vx,Baerwald:2010fk} and the AGN model introduced in \Refer~\citep{Rodrigues:2017fmu}.
  For BL Lacs, we implement a one-zone model where the cosmic rays interact with the non-thermal radiation produced in the jet; for FSRQs, additional photons emitted from the broad line region and the dust torus provide additional targets for the photohadronic interactions, which enhance the neutrino emission. For the extragalactic propagation of the UHECRs from the source to Earth, we use the novel numerical code \textsc{PriNCe} \citep{Heinze:2019jou}.
  
  
  We find that in high-luminosity sources, especially in FSRQs, the highly efficient photohadronic interactions lead to abundant neutrino production and to an extensive nuclear cascade. Low-luminosity BL Lacs, on the other hand, are highly efficient UHECR emitters and inefficient neutrino emitters because of the low photon densities in the jet, which allow the accelerated UHECRs simply to escape the source without interacting -- meaning that they exhibit a rigidity-dependent maximal energy as typically required in UHECR fits. Furthermore, the strong negative cosmological evolution of these sources also leads to minimal cosmic ray energy losses during propagation (as shown in \figu{blazardistro}, most low-luminosity BL Lacs have a redshift $z<0.5$).
  
  %

  We have tested a wide range of values of AGN properties, such as the baryonic loading, the cosmic-ray acceleration efficiency, and the size of the radiation zone (see appendix B where the AGN populations are discussed). We have found that not all of these parameters can be similar across all sources: at least the baryonic loading has to be higher for low-luminosity BL Lacs compared to high-luminosity AGN. The reason is that the efficient neutrino emission from high-luminosity jets would violate PeV-EeV neutrino bounds. This means that current neutrino observations break a possible parameter degeneracy and provide evidence that AGN of different populations must have different properties if they are to power the UHECR flux. This may also possibly point to different cosmic ray acceleration mechanisms in these two source classes.

  \begin{figure*}[tbp]
  \includegraphics[width=\linewidth]{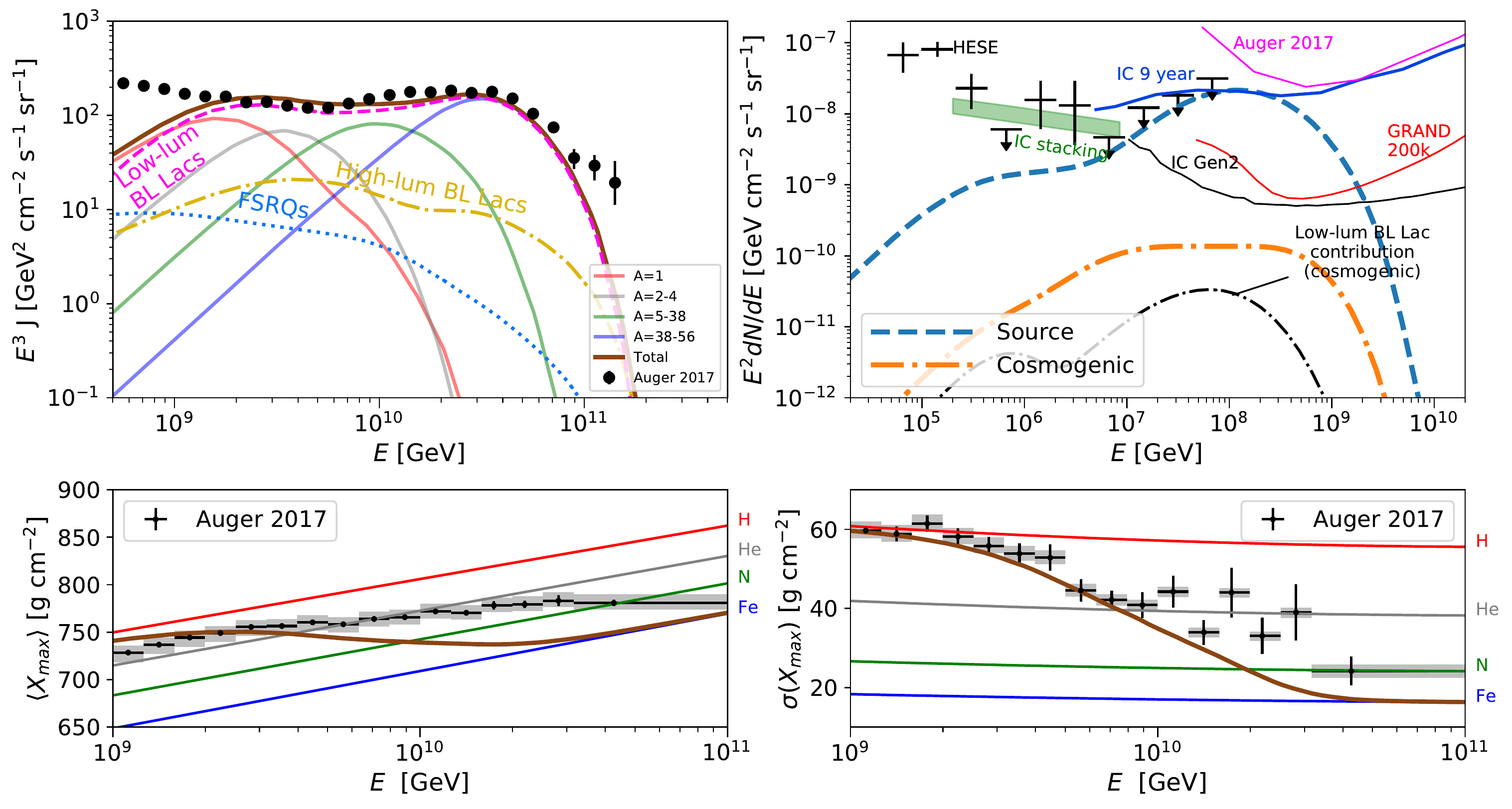}
  \caption{Description of UHECR spectrum and composition as well as predicted neutrino fluxes.
  \textit{Top left:} Simulated UHECR spectrum from the entire AGN population (dominated by low-luminosity BL Lacs), compared to data from the Pierre Auger Observatory (\textit{Auger}, \citep{Fenu:2017hlc}). \textit{Top right:} maximum (all-flavor) diffuse neutrino flux (dominated by FSRQs) that can be obtained self-consistently without violating current IceCube observations, namely the flux of HESE events (black, \citep{Aartsen:2015knd}), the stacking limit for blazars assuming a spectral index of 2.2 (green band, \citep{Aartsen:2016lir}), and the upper limits up to extremely high energies (blue curve, \citep{Aartsen:2018vtx}). Also shown are the sensitivity of \textit{Auger} (magenta, \citep{Zas:2017xdj}), of the future radio array of IceCube-Gen2 (olive green, \citep{Aartsen:2019swn}) and of the planned radio neutrino detector GRAND \citep{Alvarez-Muniz:2018bhp}. The two bottom panels show the average \textit{(bottom left)} and standard deviation \textit{(bottom right)} of the depth of the cosmic-ray shower maximum, $X_{\mathrm{max}}$, compared to \textit{Auger} measurements \citep{Bellido:2017cgf}. The colored lines correspond to the values expected for different isotopes according to the Epos-LHC air-shower model \citep{Bellido:2017cgf}.  }
  \label{fig:results_combined} 
  \end{figure*}  
  
  We summarize our main result in \figu{results_combined}. 
  In the upper left panel  we can see that it is possible to interpret the shape of the UHECR flux at and above the ankle with a dominant contribution from low-luminosity BL Lacs.
  In spite of the assumption that FSRQs have the same cosmic-ray acceleration efficiency as BL Lacs (10\%, see Tab.~\ref{tab:parameters} 1 in appendix A), their contribution is softer due to their large cosmological distances (dotted blue curve).
  
  While low-luminosity BL Lacs can explain the UHECR flux, high-energy neutrinos are efficiently produced mainly in FSRQs, which dominate the spectrum shown in the upper right panel of \figu{results_combined}. In that sense, the neutrino flux is predominantly constrained by the upper limits provided by IceCube, such as the stacking limit in the PeV range, and less so by cosmic ray data. In fact, in this model FSRQs contribute to the UHECRs flux at a level of at most 10\% at EeV energies, and the neutrino flux from FSRQs is therefore not guaranteed. However, because the UHECRs emitted by FSRQs have a high proton content, their contribution does improve the composition observables below $10^{10}$~GeV (see lower panels). In addition, if the baryonic loading of FSRQs is to be of the same order of magnitude as that of low-luminosity BL Lacs, a high neutrino flux is more naturally expected.

  While the identification of FSRQs as neutrino emitters and of BL Lacs as UHECR emitters is in agreement with the previous literature \citep{Murase:2014foa}, we additionally conclude that the neutrinos emitted by the sources can actually outshine the overall flux of cosmogenic neutrinos.
  This shows that in future searches in the EeV range, high-energy neutrinos from FSRQs should outshine the overall cosmogenic contribution from AGN, an important result for the next generation of EeV neutrino telescopes.
  For example, source neutrinos point directly to the sources, which allows for different detection techniques such as stacking searches, flare analyses or multi-messenger follow-ups. On the contrary, cosmogenic neutrinos may be isotropically distributed\footnote{In general, cosmogenic neutrinos are not necessarily isotropically distributed (see \eg\ Ref.~\citep{Essey:2010er}). However, since the \textit{Auger} results indicate that most UHECRs at the highest energies are heavy nuclei, we expect significant deflections in extragalactic magnetic fields, leading to a high level of isotropy in the cosmogenic neutrino flux.}.
  Interestingly, the same FSRQs that may dominate the EeV neutrino flux may also contribute a few events at PeV energies.
  
  Regarding the composition observables (lower panels of \figu{results_combined}), the result captures the general tendency of a heavier composition with energy. However, the predicted composition at high energies is too heavy compared to \textit{Auger} observations, because the proton-rich emission from FSRQs has a corresponding neutrino flux that is constrained by the current IceCube limits -- as shown in the upper right panel. This discrepancy may indicate that 
  additional parameters of the different AGN populations could be different, such as the initial cosmic-ray composition in the sources (which we fixed to a Galactic-like composition), or the acceleration efficiency (see Fig.~\ref{fig:best_fit} in appendix B).

  \begin{figure}[tbp]
  \includegraphics[width=\linewidth]{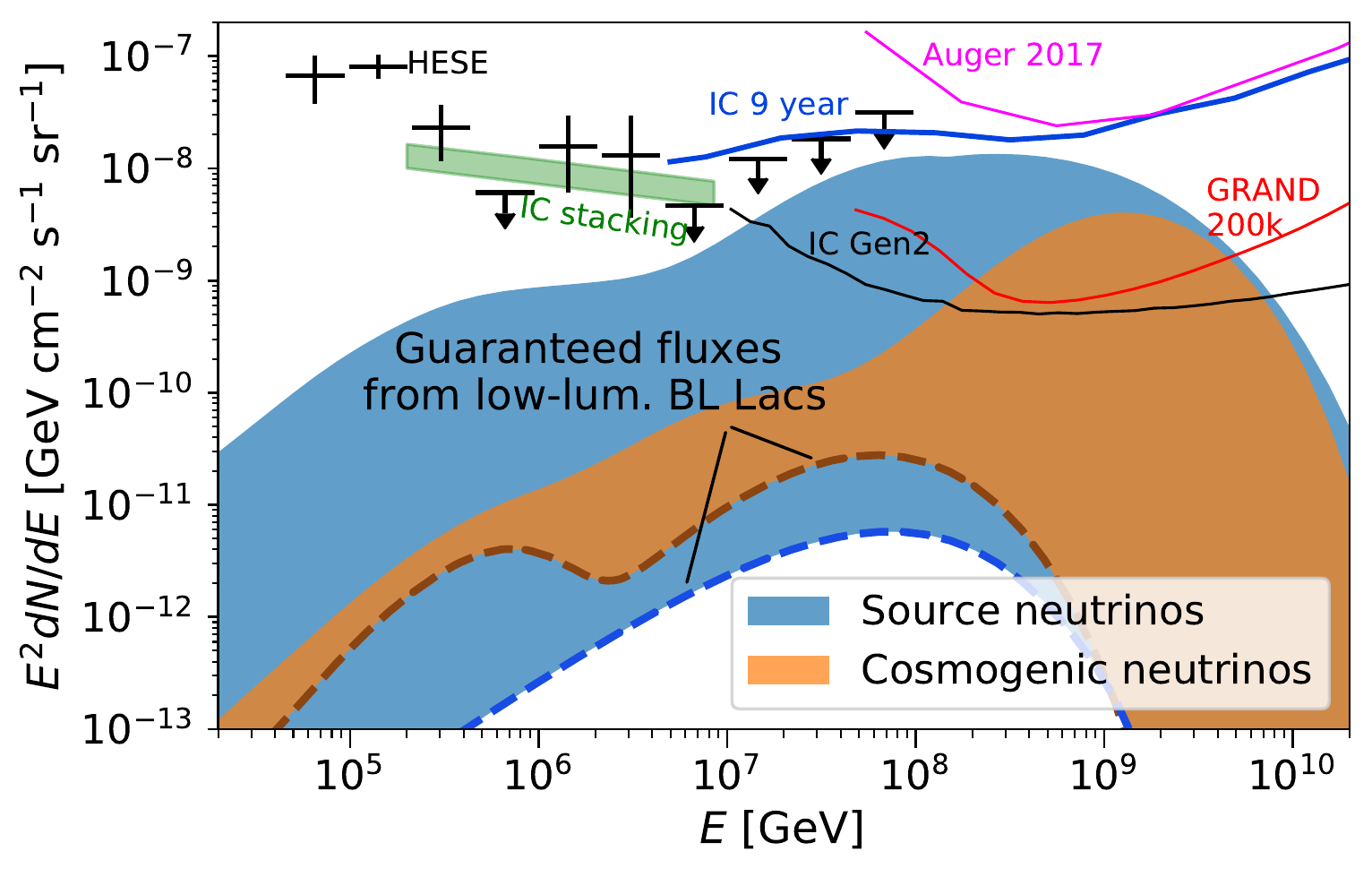}
  \caption{Predicted (all-flavor) neutrino flux range from the entire AGN population, produced through UHECR interactions inside the sources (source neutrinos, blue region) and during extragalactic propagation (cosmogenic neutrinos, orange region). The neutrino flux can saturate current IceCube limits at EeV energies while avoiding current stacking limits at sub-PeV to PeV energies. This maximum flux would originate mainly in source interactions in FSRQs, with a subdominant cosmogenic contribution. At the same time, the contribution from low-luminosity BL Lacs (dashed curves), is a guaranteed minimum flux if this source class saturates the UHECR flux as shown in \figu{results_combined}.}
  \label{fig:combined_neutrinos} 
  \end{figure}

  In \figu{combined_neutrinos} we represent the possible ranges for source neutrinos (blue band) and cosmogenic neutrinos (brown band) inferred from our analysis.
  Since the cosmic-ray acceleration efficiency of FSRQs is not constrained by UHECR arguments, the bands in \figu{combined_neutrinos} comprehend any value up to an acceleration efficiency of 100\%, in order to portray the full range of possibilities for the neutrino spectrum.
  We find that in any scenario where FSRQs dominate the neutrino flux (including the benchmark result of \figu{results_combined}), the source neutrinos dominate over the cosmogenic component. At the same time, if low-luminosity BL Lacs do indeed power the UHECRs, then the cosmogenic neutrinos from this source class constitute at least a guaranteed flux up to EeV energies (dashed curve); however, without a contribution from FSRQs, such flux would be difficult to detect with the future instruments currently proposed.

  Besides the UHECR spectrum and composition and the neutrino flux, relevant constraints to this problem can also be provided by the cosmogenic $\gamma$-ray flux and the arrival directions of the UHECRs. Our main result is expected to be fully compatible with measurements and limits on these observables. For a discussion on these topics, see appendices C and D.
  
  
  In summary, we have performed a self-consistent description of jetted AGN as the sources of the UHECRs, including a source model treating the nuclear cascade in the sources; an UHECR transport model; and a blazar population model consistent with the extragalactic $\gamma$-ray background and the evolution of the spectral energy distribution. The acceleration model and the expected injection composition have been motivated by previous results in the literature.

  We have found that low-luminosity BL Lacs can describe the shape of the UHECR spectrum and power the UHECRs, while the expected source and cosmogenic neutrino fluxes are low. In order to improve the UHECR composition observables, however, a subdominant contribution from high-luminosity AGN is required that leads to large  neutrino fluxes within the reach of upcoming experiments. We have also found evidence that the fundamental physical parameters may have to be different for the different subpopulations of AGN if this source class powers the UHECRs. A possible degeneracy in these parameters, namely in the baryonic loading, is already broken by current neutrino observations. 
  This may point towards different acceleration mechanisms at work in different AGN populations.

  Our results demonstrate that it is plausible that astrophysical source neutrinos from AGN in fact outshine the cosmogenic neutrino flux, which means that cosmogenic neutrinos could actually be the background and not the foreground at EeV neutrino energies. Since source neutrinos can be identified and disentangled with different techniques, such as stacking searches, flare analyses, or multi-messenger follow-ups, this result has profound implications for the planning and analysis of future radio-detection experiments in the EeV range, and will potentially open a new field of research. 
  An example are point-source or multiplet analyses, which may lead to the discovery of sources by finding anisotropies in the neutrino sky at the highest energies. Note that the source neutrino flux spans over many orders of magnitude in energy, and combined analysis between TeV-PeV and EeV neutrino experiments will also be of great interest. 
  
  {\bf Acknowledgments.}  The authors would like to thank Anna Franckowiak and Anna Nelles for comments on the manuscript. This project has received funding from the European Research Council (ERC) under the European Union’s Horizon 2020 research 
  and innovation program (Grant No. 646623). XR was supported by the Initiative and Networking Fund of the Helmholtz Association.
  

%

\clearpage

\appendices

\section{Details of the source-propagation model}
\label{appA}

In this section we discuss in greater detail the methods used in the main part of this letter to calculate the diffuse flux of UHECRs and neutrinos from AGN.  

\subsection*{AGN radiation model}
\label{sec:AGNradiationmodel}

The simulation of UHECR interactions in AGN jets closely follows the methods described by \citet{Rodrigues:2017fmu}.
The spectral energy distribution (SED) of each blazar depends only on its $\gamma$-ray luminosity in the \textit{Fermi}-LAT range, following  the latest parametrization of the blazar sequence \citep{Ghisellini:2017ico}, which is based on the recent \textit{Fermi} 3LAC catalog \citep{Ackermann:2015yfk}. 
The luminosity spectrum is converted into an energy density in the jet assuming that cosmic-ray interactions occur in a dissipation region modeled as a spherical blob of a given radius. These cosmic rays are assumed to be accelerated somewhere in the AGN jet, and are then injected into the blob where they interact with the photon field.

The blob is assumed to travel with a Doppler factor (from the observer's perspective) of $\delta=10$. 
The size of the blob in the co-moving frame of the jet\footnote{We represent variables given in the rest frame of the jet with a primed
symbol, and in the observer's frame as unprimed.}, $R^\prime_\mathrm{blob}=\delta \, R_\mathrm{blob}$, was fixed to 0.1~pc for all sources.

The non-thermal photon spectrum in the blob is considered to be static during the simulation, and we assume it is produced independently by a population of non-thermal electrons that are also accelerated in the jet together with the cosmic-ray nuclei.
The magnetic-field strength in the jet is assumed to scale as a power law of the $\gamma$-ray luminosity of the blazar, following Appendix~A of \Refer~\citep{Rodrigues:2017fmu}. Following closely the method in the same reference, we include the presence of external fields in FSRQs, which are reprocessed thermal radiation from the accretion disk. These include thermal emission from a dusty torus, broad-line emission from hydrogen and helium in a broad-line region (BLR) and a thermal continuum from the partial isotropization of the disk radiation. Since the size of the BLR is assumed to scale with $\gamma$-ray luminosity, for bright FSRQs the blob will be enclosed within the BLR, in which case these radiation fields will be partially boosted into the jet, where they play a major role in the photointeractions of cosmic rays.

The photohadronic interactions in the jet are calculated using the \textsc{NeuCosmA} code \citep{Hummer:2010vx,Baerwald:2010fk}, which consists of a time-dependent solver of a system of partial differential equations that describe the evolution of each particle species involved. This consists of a series of hundreds of nuclear isotopes with masses from hydrogen up to iron-56, as well as photons, pions, muons and neutrinos, which are produced through the decay of these particles. The simulated interactions include pair production, photomeson production, and photodisintegration (in the case of nuclei heavier than protons). Photodisintegration leads to the break-up of nuclear species into lighter elements, and in \textsc{NeuCosmA} this is calculated using the \textsc{Talys} model \citep{Koning:2007}.

Acceleration is not explicitly simulated. Instead, we assume that the cosmic rays are accelerated to a power-law spectrum with an exponential cutoff:
\begin{equation}
\frac{dN}{dE^\prime}\propto E^{\prime-2}\exp{\left(\frac{E^{\prime}}{E^{\prime}_{\mathrm{max}}}\right)} \, ,
\label{equ:inj_spec}
\end{equation}
where $E^\prime$ is the energy of the nucleus in the jet rest frame, $dN/dE^\prime$ is the differential energy density of this nuclear species in the jet, and $E^\prime_{\mathrm{max}}$ is the maximum injection energy where the cutoff occurs. The acceleration process is assumed to take place in an acceleration zone, before the cosmic-ray spectrum is then injected into the dissipation zone that is the blob. The maximum energy of the injected isotopes is calculated self-consistently by balancing the timescales of the acceleration process and the leading cooling process, following the method explained in \Refer~\citep{Rodrigues:2017fmu}. The acceleration timescale depends on the \textit{acceleration efficiency} parameter, $\eta_{\mathrm{acc}}\leq1$, defined as the ratio between the Larmor time of the cosmic rays and their acceleration timescale. Therefore, the value of the acceleration efficiency will determine the maximum energy $E^\prime_{\mathrm{max}}$ achieved by each nuclear species in the different sources. The acceleration efficiency was fixed to 10\% in the main result shown in \Fig~2 of the main part of this letter. This value leads to a maximum energy of iron nuclei of order 100~EeV in low-luminosity AGN, while in high-luminosity sources that value is reduced due to photohadronic energy losses.

The mechanism by which cosmic rays escape from the jet is another factor determining the emitted cosmic-ray spectrum. 
Because the transport equations depend only on energy and not on position, the escape mechanism of the cosmic rays from the jet must be introduced in the equation system as an escape rate $t^{\prime-1}_\mathrm{esc}$, which may only depend on the cosmic-ray energy.
In this study, we have assumed that the cosmic rays escape the radiation zone in the jet through Bohm-like diffusion, as discussed in \Refer~\citep{Rodrigues:2017fmu}: $t^{\prime-1}_\mathrm{esc}=cR^\prime_L(E^\prime)/R^{\prime2}$, where $R^\prime_L(E^\prime)$ is the Larmor radius of a cosmic ray with energy $E^\prime$. 
Because cosmic rays with higher energies will more easily diffuse to the edge of the radiation zone, the escape rate is proportional to the energy, leading to a relatively hard escape spectrum. A hard emission spectral index is in fact a requirement to explain the observed UHECR spectrum and composition (see e.g. \Refs~\citep{Aab:2016zth, Heinze:2019jou}).

Finally, the overall normalization of the total injection power in cosmic rays is given by the \textit{baryonic loading} of the jet, defined as the ratio between the total power in accelerated cosmic rays and the $\gamma$-ray luminosity of the source (above 100~MeV). This factor was allowed to have a different value in low-luminosity BL Lacs compared to high-luminosity BL Lacs and FSRQs. While this choice may seem purely ad hoc, it is in fact in part a result of the study itself. As can be seen in \Fig~2 of the main part of this letter, the baryonic loading of high-luminosity blazars is limited by the IceCube stacking limit. That is because most of these sources are resolved in $\gamma$-rays, and no significant statistical correlations have been found between the arrival directions of the IceCube neutrinos and the positions of these high-energy sources. As shown in \tabl{parameters}, the maximum value of the baryonic loading in FSRQs is 50. At the same time, as discussed earlier, we know that AGN jets should be capable of producing UHECRs, and can therefore contribute to the observed UHECR flux. Under the premise that AGN alone exhaust this flux, the baryonic loading of each source must be in the order of 100, as derived in previous studies (\eg~\citep{Murase:2014foa}). Therefore, it is necessarily the case that only a sub-population of low-luminosity BL Lacs, most of which unresolved in $\gamma$-rays, should have high baryonic loadings. If cosmic rays escaped the AGN jet through a more efficient mechanism, such as advection, the necessary baryonic loading of BL Lacs would be lower, but as mentioned above, such mechanism is disfavored by UHECR observations.

To obtain the combined UHECR/neutrino result shown in the main part of this letter, we scanned a range of physically allowed values of blob radius, acceleration efficiency, and baryonic loading. In each simulation, the blob radius and acceleration efficiency were considered the same across all AGN, while, for the reasons discussed above, the baryonic loading was allowed a different value for low-luminosity BL Lacs and high-luminosity blazars. In the result shown in the main part of this letter, the blob radius is $3\times10^{17}~\mathrm{cm}$, and the cosmic-ray acceleration efficiency 10\%, while the baryonic loading values are reported in \tabl{parameters}. These parameter values allow for the best description of the UHECR spectrum, while obeying current IceCube neutrino constraints and with the assumption of a Galactic-like composition of the cosmic rays in the source.

In high-luminosity FSRQs, the cosmic rays that escape the AGN jet will continue interacting with external fields of thermal and atomic broad line emission from these structures. We therefore implement a three-zone model for cosmic-ray escape in these sources. This leads to an additional cooling of the UHECRs and additional neutrino production in these bright FSRQs. The threshold above which it becomes relevant to consider these additional zones is related only to the $\gamma$-ray luminosity of the FSRQ, as detailed in \Refer~\citep{Rodrigues:2017fmu}; see also that reference for further details about the assumptions and the numerical implementation of this model.  

Although arguably the blazar sequence is not an accurate description of the variety of photon spectra among known blazars, we have numerically confirmed that the main result of this study does not depend very significantly on the shape of the non-thermal photon fields, and therefore on the adoption of the blazar sequence. The reason for this is two-fold: UHECRs are emitted mainly by low-luminosity AGN, which are optically thin to photointeractions; and neutrinos are emitted mainly by high-luminosity FSRQs, produced mainly through photomeson interactions with external thermal fields, as described above, while the non-thermal SED plays a secondary role in neutrino emission (see Refs.~\citep{Rodrigues:2017fmu, Murase:2014foa} for details). Therefore, for a different assumption on the non-thermal spectra produced in AGN jets, a fit similar to that of \Fig~2 of the main section can be found by varying the parameters of \tabl{parameters} within physically acceptable values.

\subsection*{Cosmic-ray composition and the reacceleration assumption}
\label{sec:reacceleration}

As described in the main part of this letter, we fix the composition of the injected cosmic rays to that suggested by \citet{Mbarek:2019glq}, who studied cosmic-ray reacceleration by AGN jets (see also \Refs~\citep{Kimura:2017ubz,Matthews:2018rpe}).
The reacceleation of cosmic rays up to energies of $\sim$100~EeV is an assumption, as well as a motivation, of our study. 
However, since we do not explicitly model cosmic-ray acceleration, but only their radiative interactions, it is important to clarify the consistency of this assumption.

The first thing to note is that in our model, neutrino production and UHECR emission from AGN are, to a certain extent, decoupled processes. That is because while FSRQs are efficienct neutrino emitters, the most abundant BL Lacs are in fact very dim sources, and therefore optically thin to photointeractions. In that sense, the parameters of the radiation model such as the blob size and the target photon spectrum do not affect UHECR emission from low-luminosity BL Lacs: the cosmic-ray spectrum that is accelerated is practically unchanged from acceleration until it is emitted by the source. Therefore, in these sources, the model described in the previous section is essentially reduced to cosmic-ray injection in the jet up to $\sim$100~EeV, followed by its emission. Some residual neutrino production does take place in low-luminosity BL Lacs, but only at lower energies of up to 100~PeV, as shown in \Fig~2 of the main part of this letter and as discussed also in Ref.~\citep{Palladino:2018lov}. The model is therefore compatible with the physical scenario where Galactic-like cosmic rays are reaccelerated and emitted without necessarily entering deep into the jet.

On the other hand, in FSRQs the parameters of the radiation model are in fact relevant (because they impact neutrino emission), but the reacceleration mechanism is not constrained either by composition arguments or by energetic requirements.
The first reason is that, as discussed in Ref.~\citep{Rodrigues:2017fmu}, neutrino production in the sources is largely independent of the cosmic-ray composition (except in the most powerful quasars, which are very rare), and therefore our result is not sensitive to the particular composition of UHECRs in FSRQs. Secondly, the predicted diffuse neutrino fluxes peak at most at 1~EeV (which is determined by the spectrum of the external target photons), and therefore the cosmic rays producing these neutrinos need to peak at $\sim$1~EeV only in the rest frame of the jet. Cosmic rays from outside the jet eventually reaccelerated to higher energies will not influence the results.

In summary, although acceleration is not explicitly calculated in this work, the model is consistent with the scenario of cosmic-ray reacceleration in AGN jets.

\subsection*{Extrapolation to the entire AGN population}

The cosmological evolution of blazars follows~\citet{Ajello:2011zi, Ajello:2013lka} and is described in terms of a distribution in redshift, luminosity and spectral index (assuming a power-law spectrum in the Fermi-LAT energy window). The adoption of the model by Ajello et al., which is consistent with $\gamma$-ray background observations, ensures that the present analysis also shares that consistency. We then integrate the distribution over the spectral index, obtaining a distribution in redshift and luminosity, as shown in \Fig~1 from the main part of this letter. In this description, high-luminosity BL Lacs ($L_\gamma \geq 3.5 \times 10^{45}$) and FSRQs have positive source evolutions, with a peak around redshift $z=1$. These objects are quite rare, with typical local densities $<1$~Gpc$^{-3}$.
On the other hand, low-luminosity BL Lacs ($L_\gamma < 3.5 \times 10^{45}$) have a negative evolution with redshift, which means they are most abundant in the local Universe. These objects have local densities higher than the high luminosity ones, with typical values between 1 and 100~Gpc$^{-3}$.

This blazar evolution model includes sources that fall below the \textit{Fermi}-LAT sensitivity and are therefore only theoretically expected, as shown in \Fig~1 from the main part of this letter. The negative evolution implies that there are a large number of relatively nearby BL Lacs that contribute to the UHECR spectrum at the highest energies. If we were to impose a redshift cut on to the blazar population at, say, 100~Mpc (note Mrk~421 is located at 133~Mpc), then the predicted UHECR spectrum would suffer a cutoff at $\sim3\times10^{10}~\mathrm{GeV}$ due to photointeractions during propagation. The observed spectrum below this energy could still be explained by low-luminosity BL Lacs located further away than 100~Mpc, requiring a three-fold increase in their baryonic loading compared to the result of our baseline model. Note that due to deflections in magnetic fields not only blazars can contribute to the UHECR flux, but also AGN with jets pointing in other directions.

\subsection*{Cosmological propagation of the UHECRs}

The simulation of the propagation of UHECRs from their sources to Earth is performed using the \textsc{PriNCe} code~\citep{Heinze:2019jou}. Written in \textsc{Python}, \textsc{PriNCe} uses a vectorized formulation of the UHECR transport equation taking into account the full nuclear cascade due to photodisintegration and photomeson production, as well as energy losses due to cosmological expansion and pair production. \textsc{PriNCe} has been extensively cross-checked to reproduce results from both \textsc{CRPropa}~\citep{Batista:2016yrx} and \textsc{SimProp}~\citep{Aloisio:2017iyh}. Photodisintegration interactions were calculated using the Puget-Stecker-Bredekamp (PSB) parametrization \citep{Puget:1976nz}. We adopted the Epos-LHC air-shower model \citep{Bellido:2017cgf} to convert the composition of UHECRs arriving at Earth into values for $\langle X_{\mathrm{max}}\rangle$ and $\sigma(X_{\mathrm{max}})$. Further details regarding the \textsc{PriNCe} code can be found in Appendix~A of \Refer~\citep{Heinze:2019jou}.
Evidently, the utilization of a different model would change the interpretation of the predicted UHECR composition. For example, using the Sybill~2.3 model \citep{Riehn:2015oba} would lead to larger $\langle X_{\mathrm{max}} \rangle$ and $\sigma(X_{\mathrm{max}})$ values compared to those shown in \Fig~2 of the main part of this letter. However, in general these changes can be compensated for by assuming an escape mechanism that leads to a harder UHECR spectrum, or by assuming a larger radius of the production region, thus reducing the extent of photodisintegration in the sources.

Note that unlike electromagnetic radiation and neutrinos, UHECRs typically do not point back directly to the source (due to deflections whose severity depends on composition and energy). At the same time, this also means that UHECRs escaping sideways from their sources may be scattered back into the observer's direction by magnetic fields, which implies that non-blazar AGN also contribute to the overall flux. In that sense, a study of UHECR emission from blazars must also include misaligned jetted AGN, although in observational terms these objects fall into different categories. 
In our model, it is assumed that these misaligned AGN have similar properties to blazars regarding UHECR acceleration, considering a) that the available data support a unified view of these objects and b) the Universe is isotropic, i.e., there is no reason to believe that objects pointing to Earth are special \citep{Urry:1995mg}.
This additional population is then taken into account indirectly, through the inclusion of the beaming factor of the jet as a correction to the apparent local rate of each blazar class (which is implicitly performed by including the solid angle boost when transforming the emitted flux from the blob to the black hole/source frame \citep{Rodrigues:2017fmu}). 
While there exist more generic approaches to this problem that could be independent of this assumption, such as the inclusion of an additional population of misaligned jetted AGN with a local density higher than that of blazars, our method should be in fact mathematically equivalent, leading to the same effective result.

\section{Breaking down the AGN sub-populations}

\begin{table*}[tp]
  \caption{Values of two AGN parameters: baryonic loading (total for all CR species) and acceleration efficiency. The definitions of these and other AGN parameters follow \Refer~\citep{Rodrigues:2017fmu} and are described in detail in appendix A. The parameter values indicated as \textit{main result} represent the result of \Fig~2 from the main part of this letter. There, the acceleration efficiency, as well as all the other source parameters, has been fixed for all AGN. The four middle rows refer to the single-population examples discussed in appendix B. In each of those examples, we adopt the value of baryonic loading that is necessary to reach the observed UHECR flux (at some energy bin) with only one of the three blazar sub-populations: FSRQs, high- and low-luminosity BL Lacs (HL and LL BL, respectively). We also explore different values of acceleration efficiency in order to illustrate the different properties of these populations. The final row gives the parameters for the example in appendix B where the different AGN populations are combined, while allowing for different acceleration efficiencies.}
  \label{tab:parameters}
    \begin{tabular}{llccc}
      \toprule
      Example & AGN class & \hspace{2mm}Baryonic loading\hspace{2mm} & \hspace{2mm}Acceleration efficiency \hspace{2mm}\\
      \colrule
                                                      & LL BL                                    & 380    & \multirow{3}{*}{0.1}   \\
      Main result (all AGN)                           & HL BL                                    & 50     &  \\
                                                      & FSRQs                                    & 50     &  \\
      \colrule
      Appendix B, \Fig~\ref{fig:LLBL_only}          & LL BL only                               & 400    & 0.1 \\
      Appendix B, \Fig~\ref{fig:HLBL_only}          & HL BL only (with efficient acceleration) & 40     & 0.7 \\
      Appendix B, \Fig~\ref{fig:FSRQ_only_high_eta} & FSRQs only (with efficient acceleration) & 170    & 1.0 \\
      Appendix B, \Fig~\ref{fig:FSRQ_only}          & FSRQs only (with advective CR escape)    & 20     & 0.1 \\
      \colrule
                                                      & LL BL    & 360    & 0.1 \\
      Appendix B, \Fig~\ref{fig:best_fit}           & HL BL    & 1      & 1.0 \\
                                                      & FSRQs    & 20     & 1.0 \\
      \botrule
    \end{tabular}
  \end{table*}

\begin{figure*}[tp]
  \includegraphics[width=\linewidth]{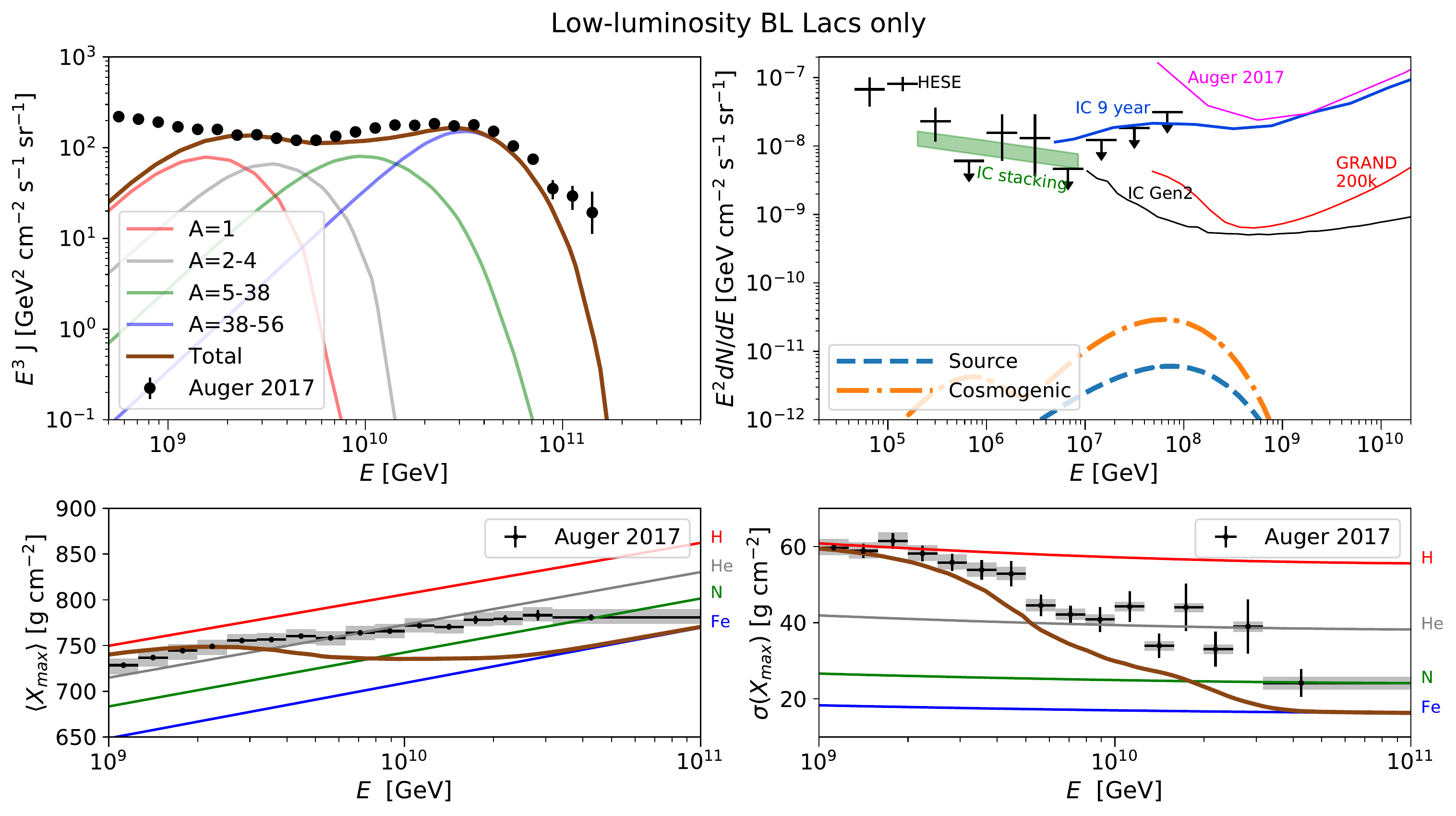}
  \caption{Expected UHECR spectrum and composition as well as source and cosmogenic neutrino fluxes for a model with only low-luminosity BL Lacs (see \tabl{parameters} for simulation parameters). Measured data, limits and expected sensitivities are the same as in \Fig~2 of the main part of this letter. This source class alone can explain the \textit{Auger} flux in a large energy range at and above the ankle. However, without the contribution from high-luminosity BL Lacs and FSRQs, the overall composition above $10^{10}$~GeV is heavier than in a combined result involving all AGN, leading to a worse description of the \textit{Auger} measurements.
  }
  \label{fig:LLBL_only}
\end{figure*}

\begin{figure*}[tp]
  \includegraphics[width=\linewidth]{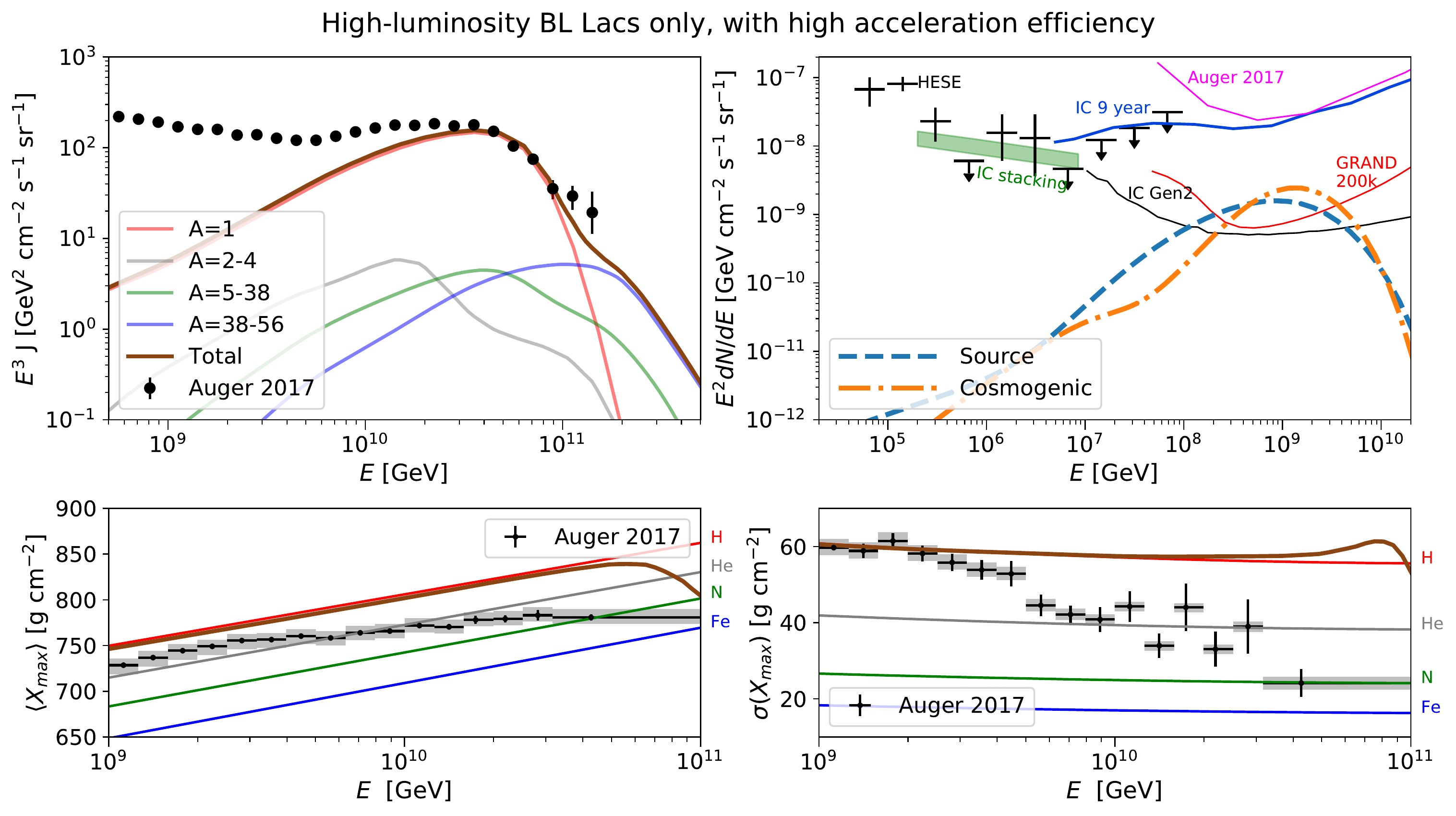}
  \caption{Expected UHECR spectrum and composition as well as source and cosmogenic neutrino fluxes for a model with only high-luminosity BL Lacs (see \tabl{parameters} for simulation parameters). With these parameter values, this source class alone can exhaust the \textit{Auger} flux above the ankle, while simultaneously emitting a flux of cosmogenic neutrinos in the EeV range that is above the planned sensitivity of IceCube Gen2 and GRAND 200k. However, the emitted UHECR composition is strongly mixed and dominated by light nuclei up to $\sim 10^{11}$~GeV, due to the high luminosity and the advective escape mechanism used in this model. This therefore gives an UHECR composition at Earth which is not compatible with the measured $\sigma( X_{\mathrm{max}} )$ data by \textit{Auger}.}
  \label{fig:HLBL_only}
\end{figure*}

\begin{figure*}[tp]
  \includegraphics[width=\linewidth]{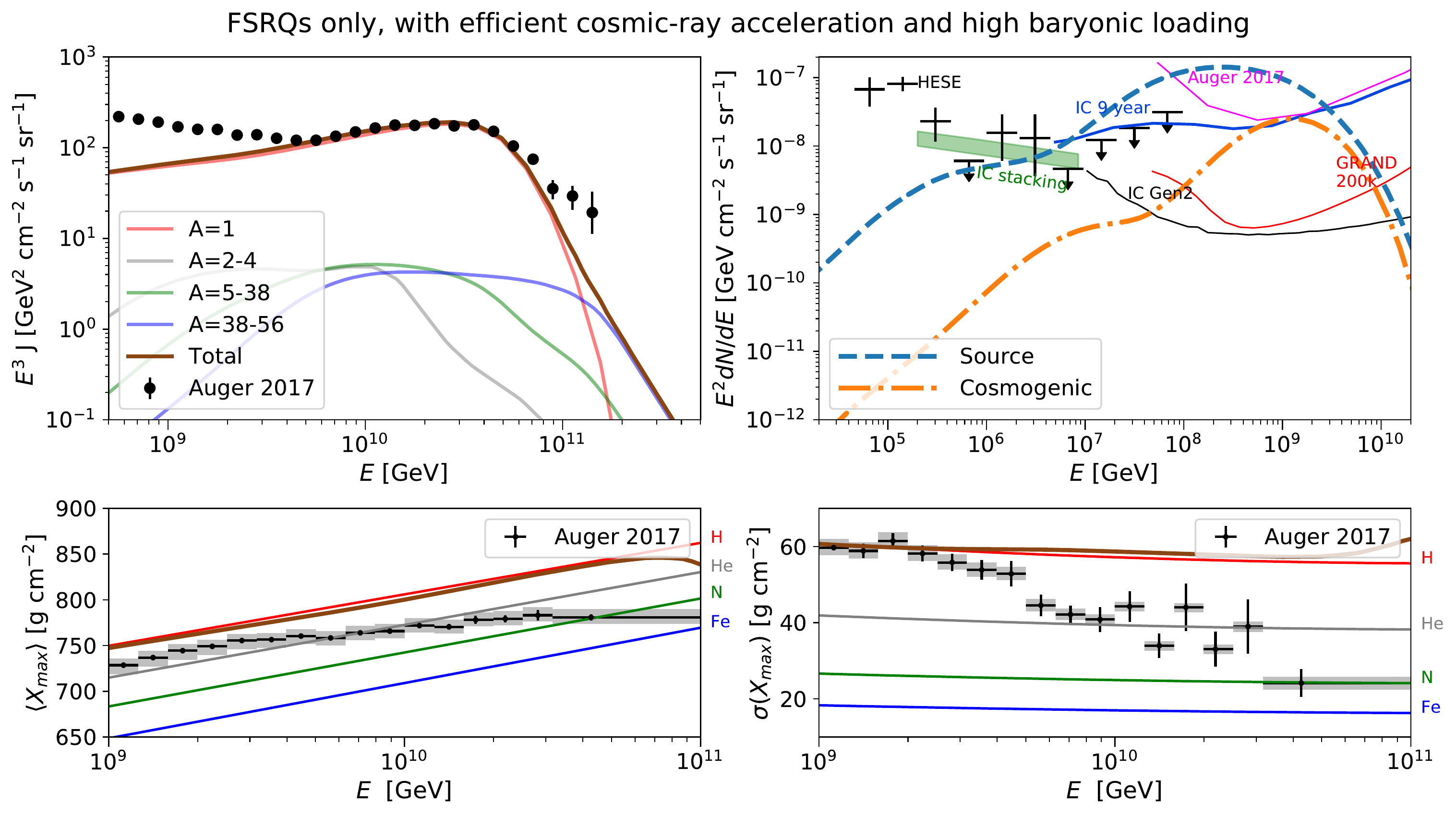}
  \caption{Expected UHECR spectrum and composition as well as source and cosmogenic neutrino fluxes for a model with only FSRQs (see \tabl{parameters} for source parameters). Here we consider  the case where FSRQs have a high acceleration efficiency. Unlike the main result of the letter, in this case the cosmic rays emitted by FSRQs peak in the ultra-high-energy regime. If we were to demand that this population exhaust the observed UHECR flux at the peak, the current neutrino limits and observations from IceCube would be violated, due to the high neutrino efficiency of these sources.
  } 
  \label{fig:FSRQ_only_high_eta}
\end{figure*}

\begin{figure*}[tp]
  \includegraphics[width=\linewidth]{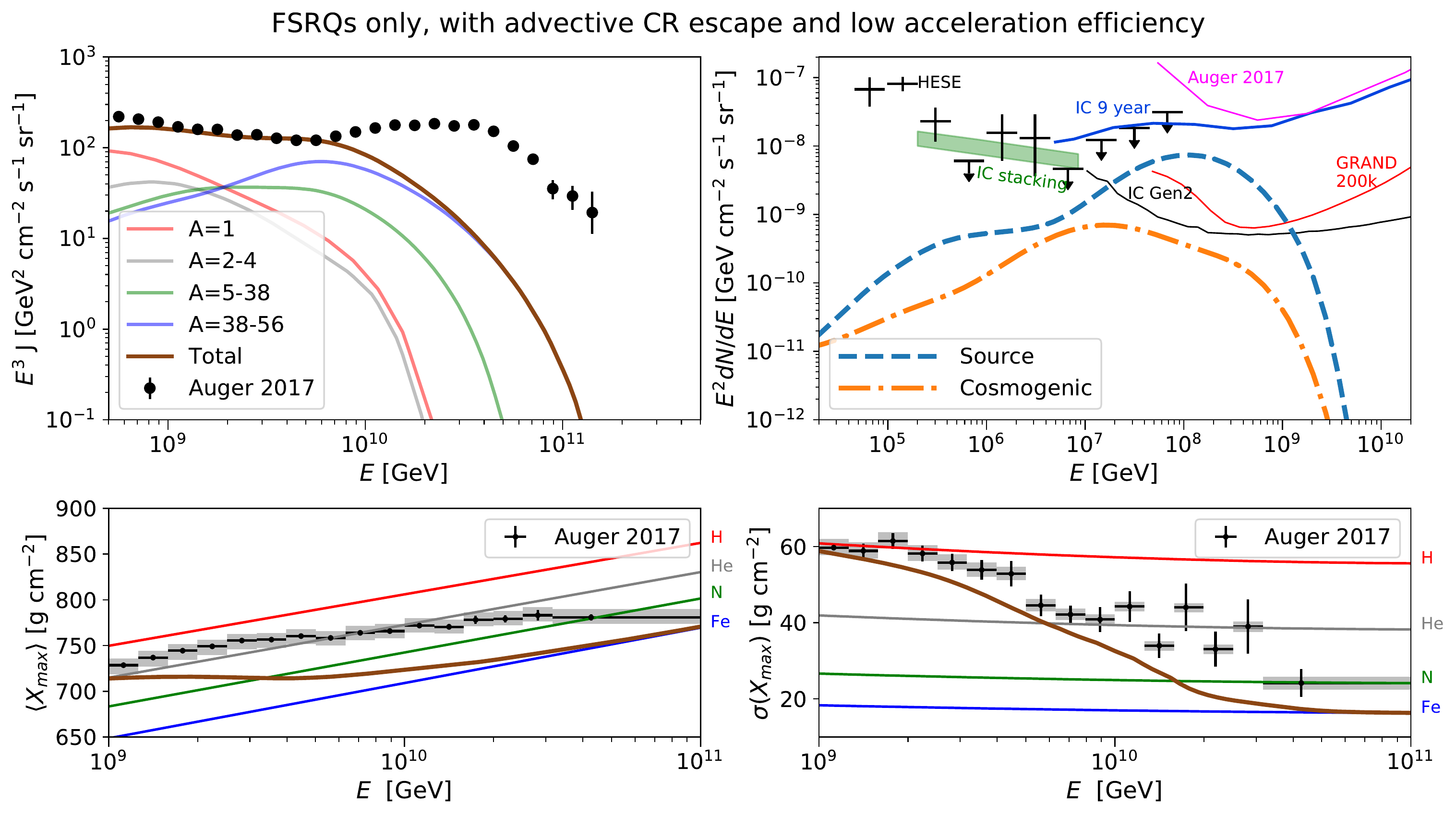}
  \caption{Expected UHECR spectrum and composition as well as source and cosmogenic neutrino fluxes for a model with only FSRQs (see \tabl{parameters} for simulation parameters). Since advection is a more efficient escape mechanism compared to diffusion, this makes the sources better UHECR emitters, and they are therefore capable of reaching the \textit{Auger} flux level in this range without violating current IceCube limits on the neutrino flux.
  }
  \label{fig:FSRQ_only}
\end{figure*}

Here, we explore the differences between the three blazar sub-classes regarding cosmic-ray and neutrino emission. From a purely quantitative perspective, all three blazar sub-classes are capable of individually exhausting the UHECR flux in a given energy range, as long as an appropriate value of the baryonic loading is assumed. Depending on the source population, this will lead to a different corresponding spectrum of source and cosmogenic neutrinos. Furthermore, in the combined result discussed in the main part of this letter, the acceleration efficiency is fixed for all sources to 10\%. This is the value that allows for the best description of the UHECR spectrum, which is dominated by low-luminosity BL Lacs. A value higher than this would lead to the overall UHECR spectrum peaking at too high energies. In bright AGN, however, the maximum energy of the emitted cosmic rays is generally lower, due to strong energy losses from photohadronic interactions in the jet. Higher values of the acceleration efficiency can therefore be tested in these sources, leading to different peak energies of their emitted UHECR spectrum.

\textit{Low-luminosity BL Lacs}: We start by considering cosmic-ray and neutrino emission from low-luminosity BL Lacs only. These are defined as BL Lacs with a $\gamma$-ray luminosity $L_\gamma \leq 3.5 \times 10^{45}$ erg/s. The reason for this splitting point is related to their cosmological evolution, as discussed in Ref.~\citep{Palladino:2018lov} (see also \Fig~1 from the main part of this letter): BL Lacs below this threshold luminosity are characterized by a negative source evolution, whereas those above this threshold luminosity are characterized by a positive source evolution, similarly to FSRQs. This may point towards different characteristics of these two BL Lac classes. 

The isolated contribution from low-luminosity BL Lacs to the UHECR and neutrino fluxes is shown in \figu{LLBL_only}. As reported in \tabl{parameters}, the baryonic loading value was adjusted in order to exhaust the UHECR flux with this sub-class alone, while other parameters are the same as in our main result (\Fig~2 from the main part of this letter). As we can see, while the fit to the UHECR spectrum is similar, the composition is heavier due to the absence of the proton-rich contribution from more powerful blazars, especially FSRQs, leading to a worse fit of both $\langle X_{\mathrm{max}}\rangle$ and $\sigma(X_{\mathrm{max}})$. 

The neutrino flux from this source class is low due to the low density of photons inside the source, which makes the photohadronic interactions inefficient. For comparison, see the left panel of Fig.~15 of \Refer~\citep{Rodrigues:2017fmu}, where the efficiency in neutrino production is reported as a function of the blazar luminosity. The cosmogenic neutrino flux is relatively low as well, because of the negative redshift evolution of low-luminosity BL Lacs. While UHECRs can only reach Earth when they are produced in the local Universe, cosmogenic neutrinos can reach us from much farther away. Closer sources (i.e.\ for a negative redshift evolution) lead, therefore, to fewer cosmogenic neutrinos arriving at Earth.

\textit{High-luminosity BL Lacs}: A scenario involving only high-luminosity BL Lacs is shown in \figu{HLBL_only}. For these sources to emit cosmic rays up to the ultra-high-energy regime, a larger acceleration efficiency is necessary (0.7 compared to our baseline assumption of 0.1). This is due to the strong cooling of the cosmic rays at the highest energies, caused by efficient photohadronic interactions with the bright non-thermal photon fields inside the AGN jet. While the cosmic-ray spectrum is well described above the ankle, the composition in this case is too light at high energies. This is also due to the efficient photodisintegration of heavy nuclei inside the sources, as well as their positive redshift evolution. 

Concerning neutrinos, the expected flux is higher than that from low-luminosity BL Lacs, but lower than from FSRQs. This is because the neutrino emission is dominated by BL Lacs with $L_\gamma < 10^{46}$ erg/s. These BL Lacs are abundant but not as efficient as FSRQs in producing neutrinos. On the other hand, high-luminosity BL Lacs are more abundant at high redshifts than low-luminosity BL Lacs (see also \Fig~1 from the main part of this letter), leading to larger source and cosmogenic neutrino fluxes. Moreover, in this scenario the sources emit a substantial percentage of protons and helium nuclei at higher energies, which further increases the cosmogenic neutrino flux. Note that although  in this case the source neutrino flux is comparable to the cosmogenic flux, it may be possible to discriminate by means of flare analyses and stacking searches.

\textit{Flat-Spectrum Radio Quasars}: We now consider two alternative physical scenarios for FSRQs in which we neglect the contribution from other source classes. In the result discussed in the main text, the cosmic-ray flux from FSRQs is considerably lower than the one from low-luminosity BL Lacs. Their total contribution is therefore constrained by current neutrino flux limits. At the same time, considering the same acceleration efficiency in all AGN, the UHECR flux from FSRQs cuts off at much lower energies compared to low-luminosity BL Lacs, which is due mainly to efficient radiative energy losses in the source. In order to explain the \textit{Auger} flux above the ankle with FSRQs, we would need to consider a higher acceleration efficiency, which is the scenario shown in \figu{FSRQ_only_high_eta}. The two limitations to this kind of scenario are the corresponding UHECR composition, which would be compatible with protons only (bottom panels), and the high neutrino fluxes, which would exhaust the current IceCube upper limits and violate the constraint based on blazar stacking analyses (upper right panel).

The neutrino flux accompanying UHECR emission would be reduced in the case where cosmic rays escape efficiently from the source, \eg~through advective winds (see \Refs~\cite{Murase:2014foa, Rodrigues:2017fmu}).
In this case, all cosmic rays can free-stream out of the blob regardless of their energy, $t^{\prime-1}_{\mathrm{esc}}=c/R^\prime_{\mathrm{blob}}$.

Compared to diffusion, this mechanism is more efficient, which means that a source with the same baryonic loading will emit a higher cosmic-ray luminosity. 
In \figu{FSRQ_only} we show the UHECR and neutrino spectra resulting from these assumptions, where the acceleration efficiency is again set at only 10\% (see \tabl{parameters}). As we can see, compared to the result in Fig. 2 of the main part of this letter, the cosmic-ray spectrum from FSRQs can now exhaust the \textit{Auger} flux below the ankle without violating the neutrino constraints. However, the composition is heavy because a higher fraction of the heavy nuclei can escape before photodisintegrating.

In this scenario FSRQs would still emit a neutrino flux in the EeV range that would be detectable by future radio neutrino experiments. 


\textit{Combined result, allowing for different acceleration efficiencies}:
Finally, we demonstrate a combined result where we relax the condition that all blazar sub-classes have the same acceleration efficiency, a restriction that was present in the main result of this letter. Essentially, the cosmic-ray composition observables will be better explained by allowing a subdominant proton-rich contribution from FSRQs or high-luminosity BL Lacs that peaks above the ankle. As demonstrated in \Figs~\ref{fig:HLBL_only} and \ref{fig:FSRQ_only_high_eta}, this implies a high acceleration efficiency in these sources. At the same time, the shape of the UHECR spectrum constrains the acceleration efficiency in low-luminosity BL Lacs to 10\%.

\begin{figure*}[tp]
  \includegraphics[width=\linewidth]{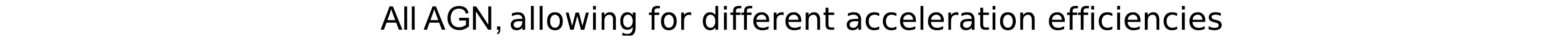}
  \includegraphics[width=\linewidth]{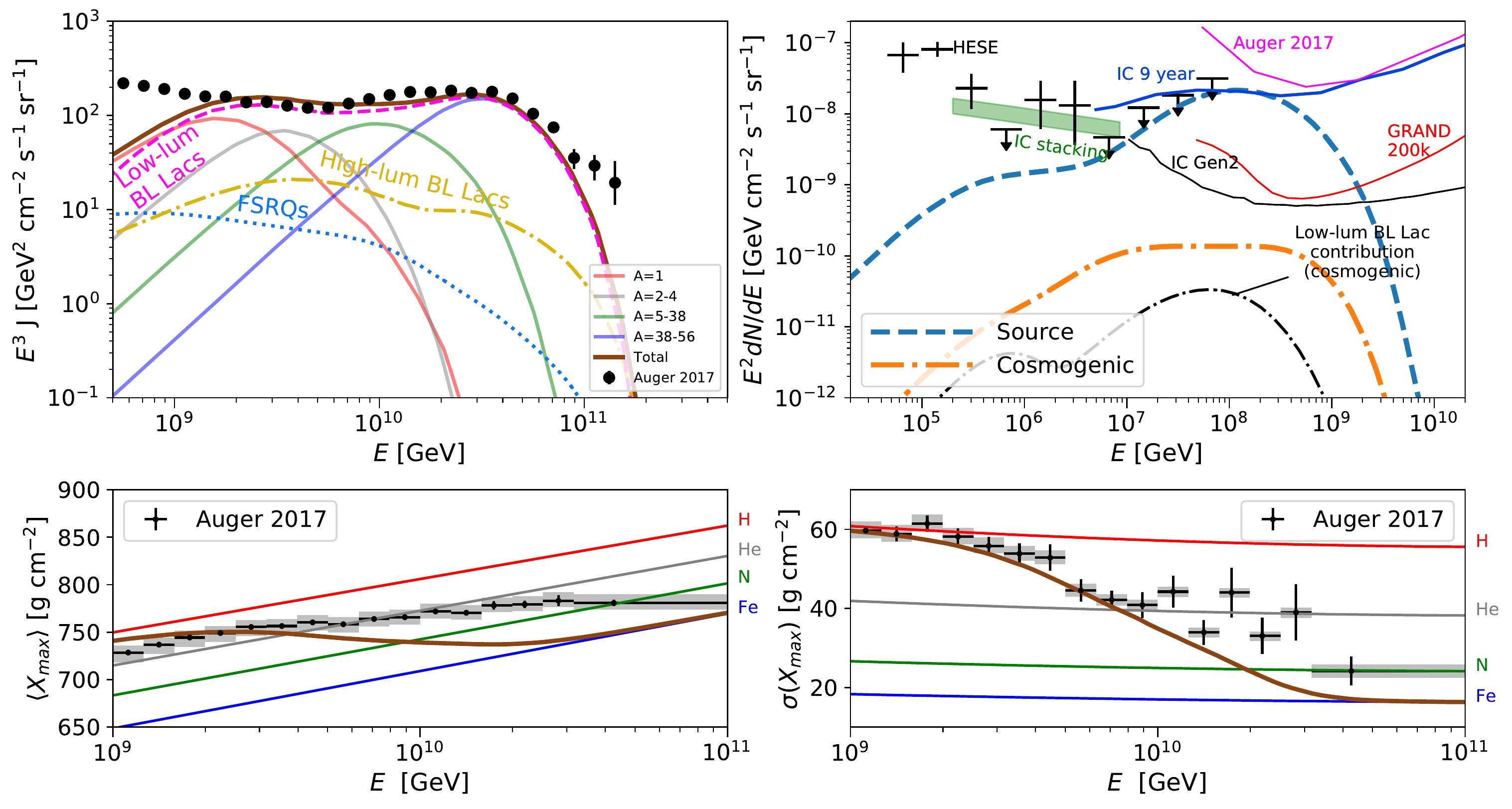}
  \caption{Expected UHECR spectrum and composition as well as source and cosmogenic neutrino fluxes from the entire AGN population, for a high acceleration efficiency in high-luminosity sources and a low acceleration efficiency in low-luminosity sources (see \tabl{parameters}). In all sources, diffusive cosmic-ray escape is considered. The high acceleration efficiency of FSRQs leads to a higher proton content up to $10^{10}$~GeV, which improves the description of the composition measurements (lower panels) compared to the result in the main part of this letter, where all AGN were assumed to have a low cosmic-ray acceleration efficiency.
  }
  \label{fig:best_fit}
\end{figure*}

In \figu{best_fit} we show the combined contribution from all AGN considering different acceleration efficiencies (\cf~\tabl{parameters}). While the spectral shape of the predicted UHECR flux is similar to the main result of this letter, the high acceleration efficiency of high-luminosity AGN allows their contribution to peak at $\sim$10~EeV. This leads to a high flux of both source and cosmogenic neutrinos, which peak at 1~EeV (upper right panel). The neutrino flux at PeV energies is unchanged, since it is not sensitive to the efficient acceleration of UHECRs to the EeV regime. At the same time, the composition above 10~EeV becomes lighter due to these sources, leading to a reduced tension with \textit{Auger} composition measurements (bottom panels).
Note that the initial composition, after acceleration and before interactions in the sources, is fixed for all AGN to the composition suggested in \Refer~\citep{Mbarek:2019glq}. Allowing for more freedom in this initial composition could further reduce the tension with the composition measurements by \textit{Auger}.

\section{Cosmogenic $\gamma$-ray fluxes}
\label{appC}

In addition to cosmogenic neutrinos, electrons, positrons and photons are also produced during the propagation of UHECRs through extragalactic space. These particles will initiate electromagnetic cascades down to GeV energies by interacting with extragalactic background light. Therefore, the measurement of the extragalactic $\gamma$-ray background (EGB) by the \textit{Fermi}-LAT telescope~\citep{TheFermi-LAT:2015ykq} could potentially provide another constraint for our models.

The diffuse cosmogenic $\gamma$-ray component depends mainly on two parameters: the cosmological evolution of UHECR sources and the composition of UHECRs emitted from those sources. The stronger the source evolution with redshift and the lighter the composition, the higher the cosmogenic $\gamma$-ray flux will be (see e.g. Refs.~\citep{Hooper:2010ze,Roulet:2012rv,Liu:2016brs,vanVliet:2016dyx,vanVliet:2017obm}). For a general scenario that fits the UHECR spectrum and composition the EGB does not provide additional constraints when error bars and model uncertainties are included~\citep{AlvesBatista:2018zui}. In addition, the same two parameters affect the expected cosmogenic neutrino flux in a similar way. Therefore, if the expected cosmogenic neutrino flux is relatively low, the expected cosmogenic $\gamma$-ray flux will be low as well. Only when the cosmogenic neutrino flux comes close to the current neutrino limits in the $10^6 - 10^8$~GeV energy range can the cosmogenic $\gamma$-ray flux be expected to give a relevant constraint (see e.g. Ref.~\citep{Supanitsky:2016gke}).

The main result, \Fig~2 from the main part of this letter, is dominated by low-luminosity BL Lacs. Since low-luminosity BL Lacs have a negative source evolution, the UHECR composition in this scenario is rather heavy and the cosmogenic neutrino flux is well below the current neutrino limits. Therefore, the EGB is not expected to significantly constrain this result. The same applies to \figu{LLBL_only} of this supplementary material (where only low-luminosity BL Lacs are considered), as well as to \figu{best_fit}. 

The case with only high-luminosity BL Lacs, \figu{HLBL_only}, has a positive source evolution and is dominated by lighter nuclei, but the expected cosmogenic neutrino flux is still well below the current neutrino limits in the range between $10^6$ and $10^8$~GeV. It is therefore unlikely that the cosmogenic $\gamma$-ray flux would be constraining. 

This is also the case for the example in \figu{FSRQ_only}: in this example, sources have a similar evolution to that of Fig.~1a of Ref.~\citep{Liu:2016brs}, and the UHECR spectrum is explained in roughly the same energy range, but the composition is heavier. The expected cosmogenic $\gamma$-ray flux should therefore be lower than in that study and will likely not provide any relevant constraint.  
In any case, this scenario can already be considered as disfavored due to the proton-like composition, in disagreement with \textit{Auger} measurements.

The cosmogenic $\gamma$-ray flux will most likely only disfavor an FSRQ-only scenario with efficient cosmic-ray acceleration and high baryonic loading, \figu{FSRQ_only_high_eta}. However, this scenario is already disfavored by the neutrino limits of IceCube and \textit{Auger} and the UHECR composition measurements of \textit{Auger}.

It is also worth noting that any prediction of the cosmogenic $\gamma$-ray flux will suffer from the additional uncertainties in the strength and shape of the extragalactic background light over a large range of redshifts (see e.g. Ref.~\citep{Berezinsky:2016jys}). As a result, the cosmogenic neutrino predictions for $E > 10^8$~GeV have in general a higher constraining power.

Considering these factors we do not expect the cosmogenic $\gamma$-ray flux to provide additional relevant constraints in any of the examples discussed in this letter. 

\section{UHECR arrival directions}
\label{appD}

In this work we focus on the UHECR spectrum and composition. Another relevant measurement is the distribution of the arrival directions of the UHECRs. In general, there is a high level of isotropy in the UHECR arrival directions; however, \textit{Auger} has recently discovered a dipolar anisotropy in the UHECR sky for $E>8$~EeV~\citep{Aab:2017tyv,Aab:2018mmi}, pointing to an extragalactic origin of UHECRs. The strength of this dipole in the UHECR sky may well be consistent with a jetted-AGN origin of UHECRs~\citep{Eichmann:2017iyr}. In addition, \textit{Auger} found an indication for anisotropies in the arrival directions of UHECRs when comparing with source catalogs of starburst galaxies and AGN for $E \gtrsim 40$~EeV~\citep{Aab:2018chp,Caccianiga:2019hlc}. Following Ref.~\citep{Caccianiga:2019hlc}, the significance of the correlation with starburst galaxies is $4.5\sigma$ and with AGN $3.9\sigma$. This is a strong indication for either of the two source candidates, but it is hardly conclusive evidence for favoring UHECR acceleration in starburst galaxies over AGN. In fact, it has been argued that the acceleration of cosmic rays up to ultra-high energies is more likely to occur in AGN vis-a-vis starburst galaxies~\citep{10.1093/mnrasl/sly099}. In particular, Centaurus A is an interesting and long-standing UHECR source candidate~\citep{Romero:1995tn,Dermer:2008cy,Biermann:2011wf,Eichmann:2017iyr}. It is the nearest FRI radio galaxy and \textit{Auger} has in fact detected a hot spot around the direction of that source.  

Whether anisotropic signals can be expected for a specific model depends, generally speaking, on the strength and correlation length of Galactic and extragalactic magnetic fields, the charge and energy of the UHECRs and the distance to and luminosity of the closest sources. If the UHECRs are mainly heavy nuclei the Galactic magnetic field alone might be strong enough to eliminate most small-scale anisotropies in UHECR arrival directions~\citep{Farrar:2017lhm}. The deflections that can be expected in extragalactic magnetic fields range from basically no deflections for protons \citep{Razzaque:2011jc,Takami:2014zva} to large deflections for iron nuclei ($> 50$ degrees) even for UHECRs with $E = 10^{11}$~GeV from sources closer than 5~Mpc~\citep{AlvesBatista:2017vob}. Besides the UHECR charge, the large uncertainties in the strength and correlation length of extragalactic magnetic fields also play a significant role.

The main result, \Fig~2 from the main part of this letter, predicts a rather heavy UHECR composition with hardly any protons at $E>10^{10}$~GeV. We therefore do not expect strong anisotropic signals in this scenario, although a weak signal from the nearest sources might still be possible depending on the extragalactic and Galactic magnetic field assumptions. The same holds for the low-luminosity BL Lacs only scenario, \figu{LLBL_only} of these supplementary materials, and for the all-AGN scenario, \figu{best_fit}. In the example of \figu{FSRQ_only}, the UHECR flux is highest in a lower energy range, for which stronger deflections in magnetic fields can be expected. We therefore do not expect any significant small-scale anisotropic signals in this case. 

On the other hand, for both the example in \figu{HLBL_only} and that in \figu{FSRQ_only_high_eta}, strong anisotropic signals could be expected due to the significant number of protons at the highest energies. This would give arrival directions comparable with the cases discussed in Ref.~\citep{Gorbunov:2002hk} or Ref.~\citep{Dermer:2008cy}. The lack of strong small-scale anisotropies in the most recent \textit{Auger} data already provides strong constraints on such scenarios.

\end{document}